\documentclass[graybox]{svmult}

\usepackage{mathptmx}       
\usepackage{helvet}         
\usepackage{courier}        
\usepackage{type1cm}        
\usepackage{makeidx}         
\usepackage{graphicx}        
\usepackage{multicol}        
\usepackage[bottom]{footmisc}
\usepackage{amssymb}
\usepackage{natbib}

\makeindex             

\newcommand\comment[1]{}

\newcommand{\wmap}{{\it WMAP}}
\newcommand{\firas}{{FIRAS}}
\newcommand{\herschel}{{\it HERSCHEL}}

\newcommand{\planck}{{\it Planck}}
\newcommand{\pixie}{{\it PIXIE}}
\newcommand{\prism}{{\it PRISM}}
\newcommand{\litebird}{{\it LiteBird}}
\newcommand{\cmbpol}{{\it CMBpol}}
\newcommand{\cobe}{{\it COBE}}
\newcommand{\groundbird}{GroundBird}

\newcommand{\ccat}{CCAT}
\newcommand{\spt}{SPT}
\newcommand{\sptsz}{SPT-SZ}
\newcommand{\simons}{Simons Array}
\newcommand{\sptnew}{SPT-3G}
\newcommand{\act}{ACT}

\newcommand{\advact}{AdvACT}

\def\n{{\bf  \hat n}}

\newcommand{\uksq}{\ensuremath{\mu {\rm K}^2}}
\newcommand{\sqdeg}{\ensuremath{{\rm deg}^2}}
\newcommand{\dksz}{\ensuremath{D_{\rm kSZ}}}

\begin{document}

\title*{Observing the Epoch of Reionization with the Cosmic Microwave Background}
\author{Christian L. Reichardt}
\institute{Christian L. Reichardt \at School of Physics, University of Melbourne, Parkville, VIC 3010, Australia, \email{christian.reichardt@unimelb.edu.au} }
%
%
\maketitle

\abstract*{
We review the observable consequences of the epoch of reionization (EoR) on the cosmic microwave background (CMB), and the resulting constraints on the EoR. 
We discuss how Thomson scattering with the free electrons produced during EoR  equates to an optical depth for CMB photons. 
The \wmap{} satellite optical depth measurement, using large-scale CMB polarization power spectra, is one of the few current constraints on the timing of cosmic reionization. 
We also present forecasts for the precision with which the optical depth will be measured by \planck{} and future satellite missions. 
Second, we consider the kinematic Sunyaev-Zel'dovich (kSZ) effect, and how the kSZ power spectrum depends on the duration of reionization.  
We review current measurements of the kSZ power and forecasts for future experiments. 
Finally, we mention proposals to look for spectral distortions in the CMB that are related to the electron temperature at EoR, and ideas to map the variations in the optical depth across the sky. 
}
\comment{
\abstract{Each chapter should be preceded by an abstract (10--15 lines long) that summarizes the content. The abstract will appear \textit{online} at \url{www.SpringerLink.com} and be available with unrestricted access. This allows unregistered users to read the abstract as a teaser for the complete chapter. As a general rule the abstracts will not appear in the printed version of your book unless it is the style of your particular book or that of the series to which your book belongs.\newline\indent
Please use the 'starred' version of the new Springer \texttt{abstract} command for typesetting the text of the online abstracts (cf. source file of this chapter template \texttt{abstract}) and include them with the source files of your manuscript. Use the plain \texttt{abstract} command if the abstract is also to appear in the printed version of the book.}
}
\section{Introduction}
\label{sec:intro}

Our earliest view of the Universe comes from the cosmic microwave background (CMB), and as such, observations of the CMB have proven to be an invaluable tool in  modern cosmology. 
CMB experiments will continue to play a vital role in testing the standard cosmological model in the future, for instance in studying inflation and neutrino physics. 
With this motivation,  CMB experiments have rapidly advanced following the first detection of temperature anisotropy using the DMR experiment on the \cobe{} satellite \citep{smoot92}, and  experimental sensitivities continue to improve  today. 
Experiments being built right now have an order of magnitude more detectors than ever before, and experiments are being proposed with more than a hundred times as many detectors \citep{snowmass13neutrinos}. 
A key distinction between the CMB and other proposed probes of reionization such as 21\,cm surveys, is that the CMB is a 2-dimensional measurement with the observables integrated along the line of sight; this rules out techniques like redshift tomography. 
Despite their 2D nature, CMB measurements have yielded two significant constraints on cosmic reionization to date.

The first major constraint on cosmic reionization came with the detection of the so-called ``reionization" bump in  large-scale CMB polarization by the \wmap{} satellite \cite{kogut03a}; the uncertainty on this measurement has decreased with each successive \wmap{} data release. 
Thomson scattering between the free electrons released by cosmic reionization  and the local CMB quadrupole produces linear polarization at the horizon scale during the epoch of reionization (EoR). 
The scattered power depends on the square of the optical depth $\tau$, so the reionization bump in polarization can break a degeneracy between $\tau$ and the amplitude of the primordial scalar perturbations $A_S$ that exists in the temperature anisotropy alone. 
The optical depth and scalar amplitude are otherwise degenerate because, for example, observing less CMB anisotropy could be explained either by increasing $\tau$ or reducing $A_s$. 
Galactic foregrounds pose a significant challenge in measuring polarization on these 10s of degree angular scales, and multiple frequencies are essential to disentangling the signals. 
Using measurements of the $\ell<20$ TE and EE\footnote{T stands for temperature anisotropy and E for polarized E-mode (zero curl) anisotropy.} power spectra from the \wmap{} satellite, \cite{bennett13} find $\tau = 0.089 \pm 0.014$. 
Data from the \planck{} satellite (and potentially other future experiments) are expected to improve upon this measurement, with the fundamental cosmic variance limit being lower by roughly a factor of seven. 
The optical depth depends on the electron number density integrated along the line of sight, and thus depends primarily on when the Universe reionizes. 

The second major CMB constraint on cosmic reionization comes from upper limits on the kinematic Sunyaev-Zel'dovich (kSZ) effect. 
Two sources contribute to the kSZ power spectrum: density variations in the late-time fully ionized Universe (homogenous kSZ) and ionization fraction variations during the EoR (patchy kSZ). 
In the standard picture of reionization, ionized bubbles form around early UV sources, with these bubbles eventually merging to form the completely ionized Universe. 
The relative velocity between these bubbles and the CMB Doppler shifts the scattered light, which  translates to a temperature shift in the CMB along that line of sight. 
The angular dependence of the kSZ power spectrum depends on the details of these bubbles, which in turn depend on the nature of the ionizing sources and the sinks of ionizing photons (i.e.~structure in the intergalactic medium) \citep{mesinger12, sobacchi14}. 
The magnitude of the kSZ power from reionization will scale with the number of bubbles, and therefore the duration of the EoR. 
The \act{} and \sptsz{} surveys have published consistent upper limits on the kSZ power \cite{addison12b,dunkley13,george14}, with the most stringent published 95\% CL upper limit on the patchy kSZ power being $\dksz < 3.3\,\uksq$ at $\ell =3000$ from the 2500\,\sqdeg{} \sptsz{} survey \cite{george14}. 
These upper limits on the kSZ power suggest cosmic reionization was not a slow process.

In Section \ref{sec:tau}, we review the physics of how the optical depth affects the CMB,  current measurements, and forecasts for future experiments. 
We do the same for the kSZ power spectrum in Section \ref{sec:ksz}. 
We review other potential observational consequences of cosmic reionization on the CMB in Section \ref{sec:other}, before concluding in Section \ref{sec:conclusion}.

\section{Mean Optical Depth}
\label{sec:tau}

\subsection{Theory}
\label{subsec:tau-theory}

\begin{figure}[t]
\includegraphics[scale=0.8]{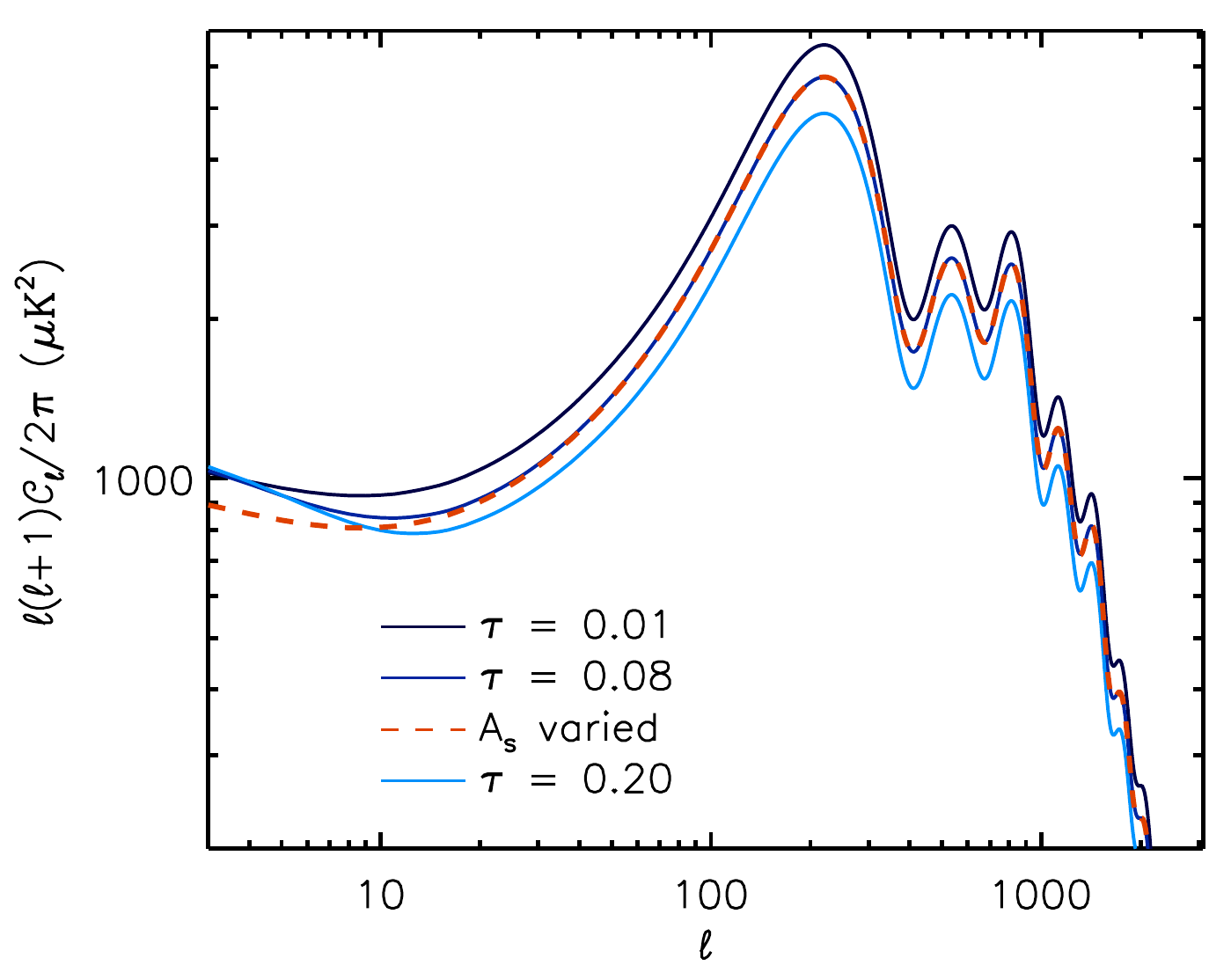}
%
%
\caption{
The optical depth $\tau$ and primordial amplitude of scalar perturbations $A_s$ have  nearly identical effects on the CMB temperature power spectrum.
The impact of increasing optical depth is shown by black to light blue lines: $\tau = 0.01, 0.08, 0.20$. 
The anisotropy power is reduced by a factor $e^{-2\tau}$ at all scales smaller than the horizon size at EoR ($\ell \gtrsim 20$). 
This reduction is nearly perfectly degenerate with a shift in the amplitude of the primordial power spectrum. 
The degeneracy in the temperature power spectrum is illustrated by the dashed red line, which mimics the dark blue $\tau=0.08$ line by reducing the primordial amplitude of scalar perturbations, $A_s$. 
}
\label{fig:tau}       
\end{figure}
\begin{figure}[t]\centering
\includegraphics[scale=0.8,clip,trim =  1.56cm  12.89cm  5.00cm  3.56cm]{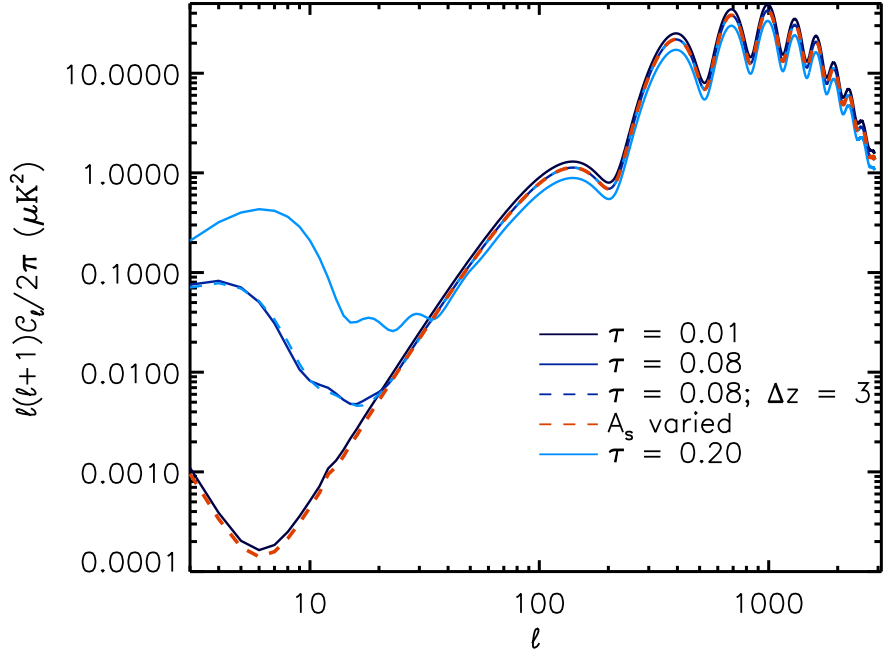}
%
%
\caption{
The impact of optical depth on the CMB $E$-mode (zero curl) polarization power spectrum. 
The solid lines show the E-mode power spectrum as the optical depth is increased: from black to light blue, $\tau = 0.01, 0.08, 0.20$. 
E-mode polarization is produced by Thomson scattering between electrons and the CMB quadrupole. 
As $\tau$ increases, the polarization power spectrum is increased on scales larger than the horizon size at EoR ($\ell \lesssim 20$). 
The induced polarized power scales as $\tau^2$, and yields the strongest constraint from the CMB on the optical depth. 
Additionally as in the temperature anisotropies, the polarization power is reduced by a factor $e^{-2\tau}$ at all scales larger than the horizon size at EoR ($\ell \gtrsim 20$). 
As in the temperature power spectrum plot, the dashed red line mimics the $\tau=0.08$ line at small scales by reducing the primordial amplitude of scalar perturbations, $A_s$ instead of increasing $\tau$. 
However, decreasing $A_s$ does not reproduce the peak at large angular scales (low $\ell$). 
Finally, the dashed blue line has the same optical depth ($\tau = 0.08$) but an EoR duration increased by 6$\times$ to $\Delta z = 3$; the $E$-mode power barely changes. 
The $E$-mode power is largely insensitive to differences between EoR scenarios that produce the same total optical depth. 
}
\label{fig:tau-ee}       
\end{figure}

The transition of the Universe from a neutral to ionized state dramatically increases the number density of free electrons that can Thomson scatter CMB photons. 
The probability that a given photon will scatter can be related to an effective optical depth to reionization:
\begin{equation}
\tau = \int n_e \sigma_T d\ell .
\end{equation}
Here, $n_e$ is the free electron number density, $\sigma_T$ is the Thompson cross-section, and the integral is over the line of sight distance. 
The integral is dominated by electrons from singly-ionized Hydrogen and Helium; doubly ionized Helium at low redshift accounts for a few percent of the total optical depth. 
Due to its integral nature, the optical depth is insensitive to the precise evolution in the ionization fraction. 
However, as the median redshift of reionization increases so will the column depth of free electrons and $\tau$. 
Therefore constraints on $\tau$ are sometimes expressed as constraints on the redshift of reionization, with only weak dependence on the assumed ionization history.

The optical depth suppresses the CMB anisotropy power at all scales smaller than the horizon size at the EoR by a factor $e^{-2\tau}$. 
This suppression is shown in Figure~\ref{fig:tau} for $\tau \in [0.01,0.20]$.\footnote{All spectra in this chapter were calculated using CAMB \cite{lewis99,howlett12}.} 
Although the magnitude of the suppression is quite large compared to measurement uncertainties, the suppression is highly degenerate with the amplitude, $A_s$, of the primordial power spectrum of scalar perturbations. 
This point is illustrated by the red, dashed line in Figure~\ref{fig:tau} which reduces $A_s$ to mimic the $\tau = 0.08$ line. 
The effect of the two parameters differs only at $\ell \lesssim 20$, and the substantial cosmic variance at these large angular scales prevents a meaningful separation with temperature data alone. 

The reionization of the Universe also creates CMB polarization. 
In general, scattered radiation from an electron in a quadrupole radiation field will be linearly polarized \cite[see e.g.,][]{rees68}. 
Thus reionization leads to linear polarization as  free electrons from reionization are exposed to the large-scale CMB quadrupole. 
The polarized signal peaks on scales larger than the horizon at EoR; the signal amplitude scales as $\tau$ and thus the power scales as $\tau^2$. 
This signal is sometimes referred to as the reionization ``bump" for reasons illustrated in Fig.~\ref{fig:tau-ee}. 
Importantly, no other parameter in the standard cosmological model produces such a feature so the parameter degeneracies are minimal. 
For better or worse,  the polarized signal encodes very little information beyond the optical depth, as demonstrated by comparing the dashed and solid blue lines in Fig.~\ref{fig:tau-ee}. 
A cosmic variance limited experiment would measure at most a couple of parameters beyond optical depth \cite{baumann09}. 
One downside is that  polarized galactic emission is major concern on these large scales (see, e.g., \cite{bennett13} for a discussion of foreground modeling). 
Finally, the number of independent modes on the sky is relatively low, which sets a fundamental cosmic variance limit on how well the amplitude can be measured, although this limit is well below current measurements.  
The reionization bump is a clean probe of the optical depth to reionization.

\subsection{Current Observations}
\label{subsec:tau-obs}

The polarization signal from reionization (``reionization bump") was first detected by looking at the temperature-polarization correlation in the first year of data from  the \wmap{} satellite \cite{kogut03a}.  
The reionization bump in the EE power spectrum was first detected in the 3-year \wmap{} bandpowers \cite{page07}. 
These measurements broke the previous degeneracy between $A_S$ and $\tau$ from the temperature data, and substantially improved cosmological constraints. 
\wmap{} is still the only experiment to have measured the reionization signal; the history of the published values is tabulated in Table~\ref{tab:tau}. 
Essentially, this is due to the large angular scales involved --- large scales strongly favor satellite experiments that cover the whole sky. 
The \wmap{} measurement is limited by a combination of instrumental noise and uncertainty in the galactic foreground modeling.

The \wmap{} polarized galactic foreground model includes two terms, dust and synchrotron (see \cite{bennett13})
The polarized synchrotron template is taken from the lowest \wmap{} frequency band (K band at 22 GHz). 
The polarized dust template starts from the model 8 dust intensity map from \cite{finkbeiner99}, with the amplitude modulated by a term to account for the magnetic field geometry, and a polarization direction taken from starlight measurements. 
The uncertainty due to the foreground modeling are on par with the statistical uncertainties. 

Recent results from the \planck{} satellite suggest that a more accurate galactic dust template reduces the inferred optical depth \citep{planck13-15, planck15-13}. 
In Appendix E,  \cite{planck13-15} find that replacing the dust template used by \wmap{} with the \planck{} 353\,GHz map changes $\tau$ by 1\,$\sigma$ to $\tau = 0.075 \pm 0.013$. 
This drop persisted through the second Planck release. 
\cite{planck15-13} found values of $\tau = 0.071 \pm 0.013$ when cleaning the large-scale WMAP9 polarization data by a polarized galactic dust template based on \planck{}'s polarized 353 GHz maps. 
Replacing the \wmap{} polarization data by \planck{} 70 GHz polarization data leads to a weaker but consistent constraint of $\tau = 0.078 \pm 0.019$. 
The joint constraint from the \wmap{} and \planck{} 70 GHz polarization data is $\tau = 0.074 \pm 0.012$. 
Setting aside all polarization data and instead using only \planck{} temperature and lensing measurements yields $\tau = 0.070 \pm 0.024$. 
These measurements are all consistent with each other and favor a lower optical depth with a midpoint to reionization around $z\simeq 9$ instead of 10.6. 
A lower optical depth would reduce the marginal tension between current CMB data and suggestions from the spectra of quasars or gamma ray bursts \cite[e.g.,][]{mortlock11, schroeder13} , and the dropoff in the luminosity function of Ly$\alpha$ emitters \cite[e.g.,][]{ouchi10, clement12} and in the Ly$\alpha$ fraction \cite[e.g.,][]{treu13, caruana14} that 
reionization ended between a redshift of 6 to 7 (see also Figure~\ref{fig:ccat}). 
We will learn more with the release of the full \planck{} polarization results (see next section).

\begin{table}[ht!]
\centering
\begin{tabular}{l|c}
\hline\hline
\rule[-2mm]{0mm}{6mm}
Source & $\tau$  \\
\hline
WMAP1 ~~&~~ $0.17 \pm 0.04$\\
WMAP3 ~~&~~ $0.089 \pm 0.030$\\
WMAP5 ~~&~~ $0.087 \pm 0.017$\\
WMAP7 ~~&~~ $0.088 \pm 0.015$\\
WMAP9 ~~&~~ $0.089\pm 0.014$\\
\hline
with Planck TT and 353\,GHz dust template: & \\
WMAP9 pol.~~&~~ $0.071\pm 0.013$\\
Planck LFI pol. ~~&~~ $0.078\pm 0.019$\\
Planck LFI pol. (no TT) ~~&~~ $0.067\pm 0.022$\\
Planck lensing + BAO ~~&~~ $0.067\pm 0.016$\\
\hline
\end{tabular}
\caption{\label{tab:tau} The optical depth measured by \wmap{} and \planck{} across data releases. 
\wmap's measurement of the optical depth has been relatively static after the second \wmap{} data release \cite{page07}. 
Differences between the first and second data release include: (1) $\tau$ is derived primarily from $EE$ instead of $TE$, (2) three times the data volume, (3) a new foreground treatment, and (4) an updated polarization analysis. 
An updated polarized galactic dust template based on the 353\,GHz channel of \planck/HFI reduces the optical depth by $\sim$\,1\,$\sigma$ (bottom half of table). 
The low-$\ell$ polarization data from 70\,GHz in Planck favors an even lower optical depth; a value that is consistent with estimates that avoid using the large-scale polarization data at all.
Future releases of the \planck{} satellite are expected to achieve $\sigma(\tau) \sim 0.005$ and the cosmic variance limit is  $\sigma(\tau) \sim 0.002$. 
} 
\end{table}

\subsection{Future Observations}
\label{subsec:tau-forecast}

Optical depth constraints from the full polarization analysis of the \planck{} satellite are expected to be released towards the end of 2015. 
Based on sensitivities from the \planck{} bluebook \cite{planck06}, the \planck{} polarized noise should be substantially lower than \wmap{}. 
Optical depth forecasts are challenging beause the \planck{} constraint is limited by the foreground subtraction rather than instrumental noise. 
However, \cite{planck06} predict $1\,\sigma$ error bars of $\sigma(\tau) \simeq 0.005$.  

A number of proposed satellite experiments hope to improve upon the \planck{} result and reach the fundamental cosmic variance limit at $\sigma(\tau) \simeq 0.002$. 
These include \cmbpol{} \cite{baumann09}, \litebird{} \cite{hazumi12}, and \pixie{} \cite{kogut11}. 
There are also a handful of ground- or balloon-based experiments, such as \groundbird{} \cite{tajima12}, that might be able  to measure these large scales.

\section{Kinematic Sunyaev-Zel'dovich effect}
\label{sec:ksz}

\subsection{Theory and modeling}
\label{subsec:ksz-theory}

The second observational signature of the EoR on the CMB is the kinematic Sunyaev-Zel'dovich (kSZ) effect.
The bulk velocity of free electrons  relative to the CMB will  introduce a Doppler shift to the scattered photons, an effect known of the kSZ effect \cite{sunyaev72, phillips95, birkinshaw99, carlstrom02}. 
In the non-relativistic limit, the kSZ effect slightly changes the observed CMB temperature,   with the temperature shift  scaling  as $(v/c) n_e$ where $v$ is the line-of-sight bulk velocity of electrons, $c$ is the speed of light and $n_e$ is the density of free electrons. 
The result is a hot spot  if the ionized gas is moving towards the observer and a cold spot if moving away.

The total kSZ signal along a line of sight  is: 
\begin{equation}
\frac{\Delta T_{\rm kSZ}}{T_{CMB}} (\n) = \sigma_T\overline{n}_{e,0}
\int d \eta a^{-2} e^{-\tau(\eta)} \bar x_e(\eta) (1+\delta_x) (1+\delta_{\rm b}) (-\n \cdot \mathbf{v}) \;,
\label{eq:ksz}
\end{equation}
where $\sigma_T$ is Thomson scattering cross-section,  $\tau(\eta)$ is
the optical depth from the observer to conformal time $\eta$, $\bar x_e(\eta)$ is the mean ionization fraction at $\eta$, $a$ is the scale factor at $\eta$,  and 
$\bar{n}_{e,0}$ is the mean electron density of the universe today. 
Perturbations in the baryon density and ionization fraction are marked by $\delta_{\rm b}$ and $\delta_{\rm x}$ respectively. 
Finally, $\n$ is the line of sight unit vector and  ${\bf v}$ represents the peculiar velocity of free electrons at $\eta$. 
A net kSZ signal thus requires perturbations in the free electron number density that are correlated with the large-scale velocity field.

The kSZ signal is naturally divided into two components. 
The homogenous kSZ signal is sourced by perturbations in the density ($\delta_{\rm b}$) of the fully ionized Universe. 
The patchy kSZ signal is sourced by perturbations in the ionization fraction ($\delta_x$) during reionization. 
Ionized bubbles are expected to form around the first stars, galaxies, and quasars. 
These bubbles eventually overlap and merge, leading to a fully ionized universe. 
The proper motion of an ionized bubble generates angular anisotropy through the kSZ effect. 
These bubbles are generically correlated with the velocity field because the ionizing sources are biased tracers of the matter distribution. 
The velocity dependence also means that larger simulation volumes are required to properly estimate the sample variance; the auto-correlation length of the velocity field is order 100 Mpc whereas that of the reionization field is order 10 Mpc. 
The patchy and homogenous kSZ components are expected to have comparable power. 

The amplitude of this patchy kSZ power depends primarily on the duration of reionization, while its shape depends on the distribution of bubble sizes. 
Both features also depend more weakly on the average redshift of reionization \cite{gruzinov98,knox98,santos03,zahn05,mcquinn05,iliev06,zahn12, mesinger12, battaglia13,calabrese14}. 
The qualitative behavior of these dependencies can understood simply: the power is linearly proportional to the number of bubbles along the line of sight which scales  with the duration. 
Similarly if the bubbles are larger, the kSZ power will peak at larger scales and vice versa. 
If cosmic reionization occurs at earlier times, the Universe is denser and the same duration leads to more kSZ power. 

There are two potentially important qualifications to this picture. 
First, the signal depends on reionization being inhomogenous. 
Perfectly homogenous sources can reionize the Universe without producing any kSZ power. 
However, in practice, this is unlikely to be significant. 
Recent work by \cite{mesinger13} shows that reionization by ultra-hard x-rays (which have corresponding long mean free paths) reduce the kSZ power by less than 0.5\,\uksq. 
Second, observations typically probe a specific angular scale, and therefore a particular weighting of bubble sizes. 
Simulations that change the angular shape, whether by changing the mean free path of ionizing photons and thus bubble sizes \cite{mesinger12} or suppressing bubbles of the relevant size with a self-regulation mechanism \cite{park13}, result in a more complicated relationship between the midpoint and duration of EoR than found by \cite{zahn12} and others.

Current data can not distinguish between the homogenous and patchy kSZ components  because both components have the same spectral dependence and similar angular dependencies.
Thus the EoR inferences depend on  accurately modeling (and subtracting) the homogenous contribution. 
The homogenous kSZ power spectrum has been simulated by a number of authors \cite[e.g.,][]{trac11,shaw12}. 
Recent predictions for the homogenous kSZ power  at $\ell = 3000$ range from 2.2 to 3.2 \uksq{} for a common cosmology, a peak-peak range of $\pm$20\%. 
This modeling uncertainty is subdominant to  current statistical uncertainties, although this is likely to change with upcoming experiments. 
Future experiments might be able to separate the components using the angular dependence (although this will be challenging), higher-order moments of the map, or cross-correlation with other observables \cite{shaw12}.

\subsection{Current Data}
\label{subsec:ksz-obs}

\begin{figure}[t]\centering
\includegraphics[width=0.43\textwidth,clip,trim =  2.97cm  12.96cm  5.09cm  3.71cm]{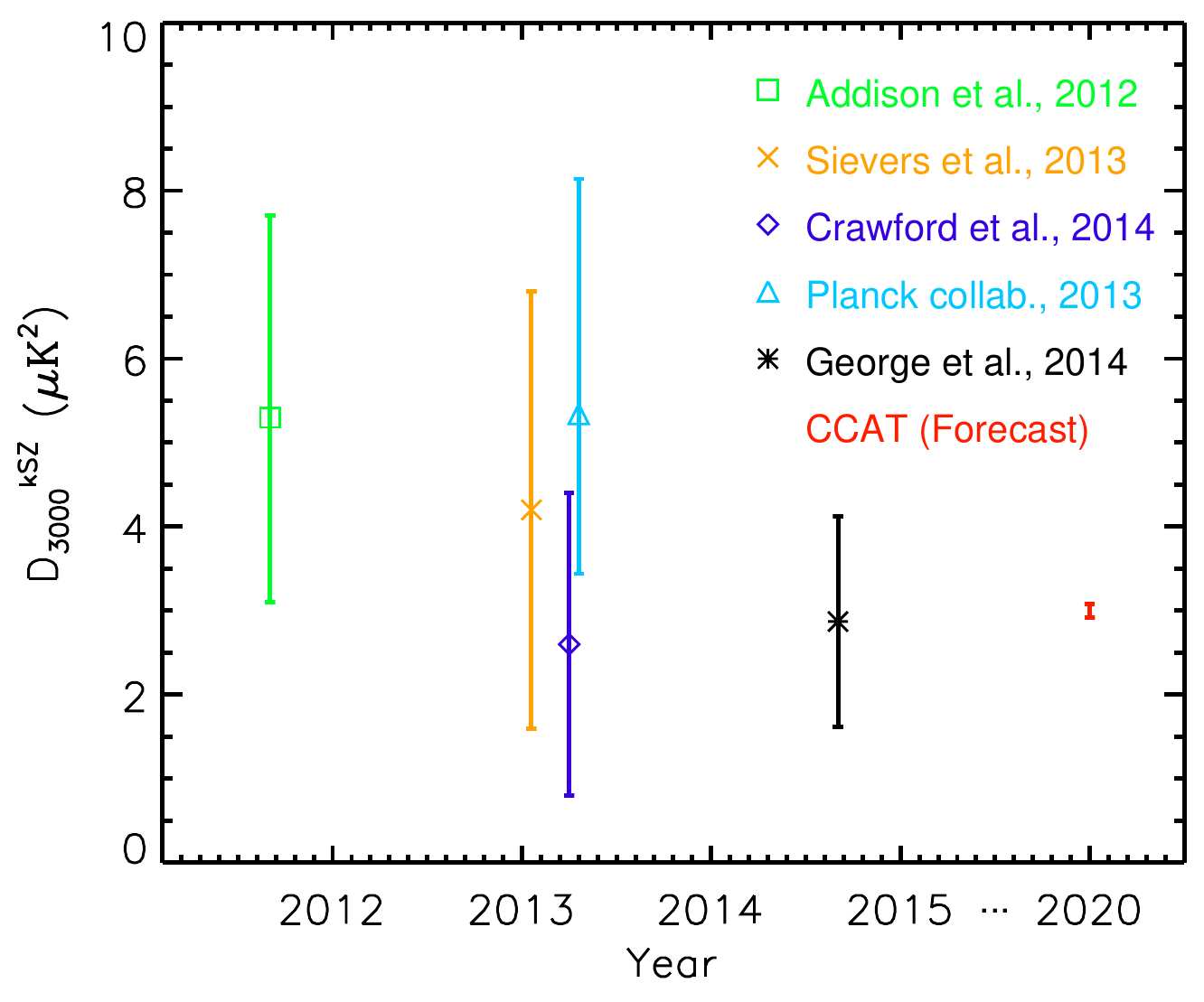}
\includegraphics[width=0.47\textwidth,clip,trim =  0.76cm  7.55cm  1.82cm  8.1cm]{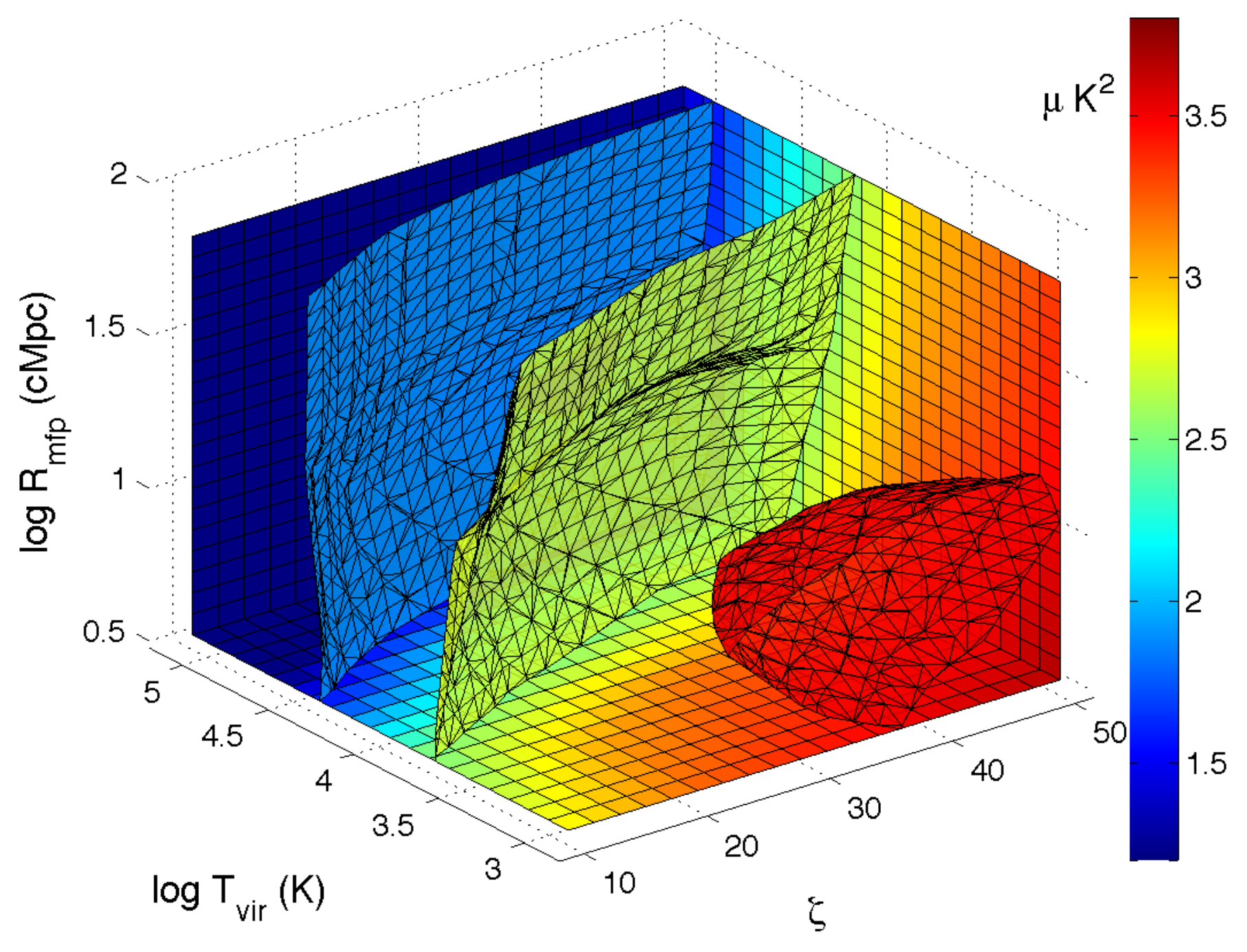}

%
%
\caption{
\textbf{Left panel:} 
Published $1\,\sigma$ constraints on the total kSZ power at $\ell = 3000$ ordered by arxiv release date. 
The most recent two constraints  using data from \act{} (green square \cite{addison12b} and orange x \cite{sievers13})
 and \sptsz{} (purple diamond \cite{crawford14} and black asterisk \cite{george14}) are shown. 
The \planck{} result (light blue triangle \cite{planck13-16}) uses both \act{} and \sptsz{} data. 
Also shown in red is a forecast  (arbitrarily centered at 3\,\uksq{}) for the proposed LWCAM instrument on \ccat{} which should begin taking data by 2020. 
Upcoming experiments like  \ccat{}, \advact{}, and \sptnew{}  should dramatically improve measurements of the kSZ power spectrum. 
\textbf{Right panel:}  
The patchy kSZ power as a function of three parameters describing EoR: the mean free path of ionizing photons $R_{MFP}$, the ionizing efficiency $\eta$, and the minimum virial temperature $T_{vir}$ of halos contributing to reionization. 
Detailed measurements of the kSZ power spectrum will place constraints on the astrophysical processes reionizing the Universe. 
\textit{Figure 6 in \cite{mesinger12}; used with permission.}
}
\label{fig:currentksz}       
\end{figure}

Figure \ref{fig:currentksz} shows current measurements of the kSZ power at $\ell = 3000$ using data from the \planck{}, \act{} and \sptsz{} experiments \cite{addison12b,dunkley13,planck13-16,george14}. 
 Using the 2500\,\sqdeg{} \sptsz{} survey, \cite{george14} report the kSZ power at $\ell=3000$ to be $\dksz = 2.9 \pm 1.3\, \uksq$. 
The measured power is consistent between experiments. 
Since the homogenous signal is expected to be above 2.0~\uksq{} \cite{shaw12}, the observed kSZ power leaves little room for patchy kSZ power . 
\cite{george14} combine these facts to set an 95\% confidence level upper limit on the patchy kSZ power  at $\ell=3000$  of $\dksz^{patchy} < 3.3~\uksq$. 
The measured kSZ power (from \sptsz) and optical depth (from \wmap) is used to importance sample a suite of reionization models as described by \cite{zahn12}, and infer constraints on the EoR. 
Defining the duration of reionization as the time the Universe takes to go from a 20\% to 99\% volume-averaged ionization fraction, the patchy kSZ limit translates to a 95\% confidence level upper limit\footnote{This limit does not include modeling uncertainties from reionization scenarios not included in simulation suite used by   \cite{zahn12}.} on the duration  of $\Delta z < 5.4$ \cite{george14}.

The power at these few-arcminute angular scales is a combination of the thermal and kinematic SZ effects, radio galaxies and dusty galaxies. 
In principle these components can be separated cleanly based on each one's unique frequency dependence. 
However in practice substantial degeneracies remain between the measured thermal and kinematic SZ powers because of the limited frequency coverage. 
The degeneracy in current data between the two SZ effects can be partially broken by considering higher order moments of the map,\footnote{Higher order moments can break the degeneracy because the degree of non-Gaussianity varies between components.} as is done in \cite{george14}, or by including additional data at other frequencies \cite{addison12b}. 
Current kSZ measurements are limited by the ability to separate these different signals.

\subsection{Future Observations}
\label{subsec:ksz-forecast}

In the near term, there is the potential to combine multi-frequency data from several experiments to better measure the kSZ power. 
The \planck{} satellite has multi-frequency coverage, but its coarse angular resolution makes it difficult to access the relevant small angular scales. 
Another recent satellite, \herschel{}, has made confusion-noise-limited maps of the cosmic infrared background (CIB) across hundreds of square degrees at frequencies 8-16$\times$ higher than the \sptsz{} observing bands. 
Modulo modeling uncertainties in understanding the CIB over such a wide frequency range, the \herschel{} data could be used to constrain the CIB and break the degeneracy between the tSZ-CIB correlation and kSZ that limits  current measurements. 
Fisher matrix forecasts predict that adding \herschel{} data to the \sptsz{} data would improve current constraints by a factor of four.

In the medium term ($\sim$2020), the next generation of experiments on the \act{} and \spt{} telescopes should lead to a substantial improvement in the kSZ constraints. 
These experiments will have dramatically lower noise levels and better frequency coverage, which will break the degeneracy between the two SZ effects in current data. 
Forecasts for these experiments are likely to be less sensitive to the CIB modeling than the \herschel{} predictions since the data is coming from a comparatively narrow frequency range near the peak of the CMB black body. 
Uncertainties from \sptnew{} and \advact{} are likely to be comparable, with the \sptnew{} survey forecast to precisely measure the kSZ power with $\sigma(kSZ) = 0.15\,\uksq$.  

\begin{figure}[t]\centering
\includegraphics[width=0.95\textwidth]{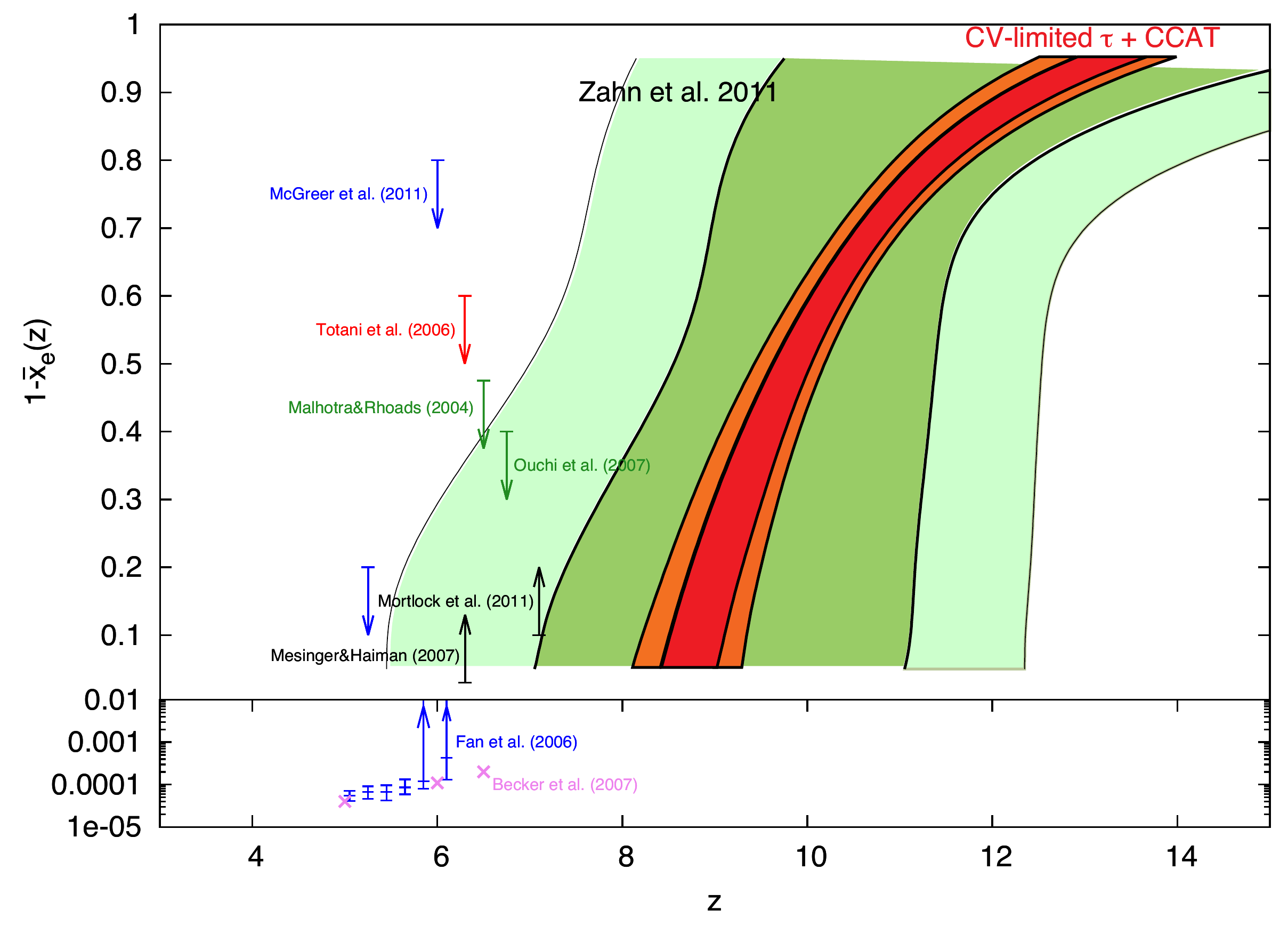}
\caption{
CMB-derived constraints on the redshift evolution of the mean neutral fraction.
The \sptsz{}+\wmap{} 68/95\% confidence ranges  for the conservative case in \cite{zahn12} are indicated by the dark/light green shading. 
Forecast constraints from \ccat{} plus a future cosmic variance limited measurement of the optical depth are shown by the red/orange shading. 
A sampling of other constraints on the neutral fraction based on quasar spectra (blue and violet constraints as well as black lower limits), a gamma ray burst (red upper limit), and Ly$\alpha$ emitters (green upper limits) is also shown. 
\textit{Image Credit: Oliver Zahn}
}
\label{fig:ccat}       
\end{figure}

In the longer term (mid-2020s), results from the first large-area surveys with the planned CCAT telescope should be published. 
Forecasts for the proposed LWCAM instrument on CCAT are shown in Figure~\ref{fig:ccat} for six frequency bands from 95 to 400\,GHz. 
CCAT could measure the kSZ power at multiple angular scales and test the angular dependence of the kSZ power spectrum. 
The shape of the power spectrum encodes information on the sources and sinks of ionizing photons.  
Work by \cite{mesinger12} has shown, for instance, that slope of the kSZ power spectrum around $\ell = 3000$ can inform us about the number density of Lyman Limit systems at high redshift. 
CCAT would also improve the measurement of the overall amplitude by another factor of two. 

\section{Other Observables}
\label{sec:other}

We finally turn to two other potentially observable signatures of cosmic reionization in the CMB. 
First, the ionized gas will distort the CMB black-body spectrum, with the magnitude of the spectral distortion depending on the gas temperature. 
Second, variations in optical depth across the sky will introduce a non-Gaussian signal in the CMB temperature and polarization anisotropies. 
These signals have yet to be detected.

\subsubsection{Spectral distortions}
\label{subsec:distort}

In Section~\ref{sec:tau}, we described the dominant effects of Thomson scattering between the CMB and free electrons during the EoR: damping of anisotropy and the polarized reionization bump. 
Beyond these effects, the photon-electron temperature difference  very slightly distorts the black body spectrum of the scattered CMB photons. 
This Compton y-distortion has a spectral dependence that can be parameterized as $\Delta I_\nu = y Y_{SZ}(\nu)$ \cite{zeldovich69}; y-distortions are commonly encountered in galaxy clusters  with the thermal Sunyaev-Zel'dovich effect  \cite{sunyaev72}. 
The amplitude y can be expressed as:
\begin{equation}
y = \int \frac{k[T_e(z) - T_{CMB}(z)]}{m_e c^2 } n_e(z) \sigma_T c dz,
\end{equation}
where k is Boltzmann constant, c is the speed of light, $\sigma_T$ is the Thomson cross-section, the quantities subscripted by $e$ are the electron temperature, mass and number density, and $T_{CMB}$ is the CMB temperature. 
The magnitude scales with the optical depth, $\tau$,  and electron temperature, $T_e$,   as \cite{kogut11}: 
\begin{equation}
 y  \simeq 1.8 \times 10^{-7} \left( \frac{\tau}{0.09} \right) \left( \frac{T_e}{\rm 10^4 K} \right). 
\end{equation}

This amplitude  is two orders of magnitude below the best current limit of $y < 1.5 \times 10^{-5}$  at 95\% confidence from the \firas{} experiment \cite{fixsen96a}. 
However satellite experiments have been proposed recently to improve the \firas{} measurement 10,000-fold (e.g.,~\pixie{} \cite{kogut11}, \prism{} \cite{andre14}). 
For instance, \pixie{} \cite{kogut11} is designed to achieve $\sigma_y \sim 10^{-9}$, and  thus would present the intriguing possibility of measuring the electron temperature during EoR at the 5\% level.\footnote{Note that there would be a degeneracy between the y-distortion induced by EoR and certain classes of alternative models that inject energy into the early Universe \cite{chluba14}. } 
The gas temperature during EoR is a function of the spectrum of the ionizing sources; measuring this temperature could distinguish between, e.g., Pop II stars and black holes \cite{kogut11}.

\subsubsection{Variations in optical depth}
\label{subsec:tauvariance}

The inhomogeneity of cosmic reionization means that there will be small variations in the optical depth along different line of sights. 
These variations imprint a non-Gaussian signal on the temperature and polarization anisotropies, in particular correlating the polarized E and B-modes. 
The reionization signal peaks on large scales ($\ell \sim 400$). 
Properly designed estimators can leverage these correlations to reconstruct the optical depth as a function of position on the sky \cite{dvorkin09, natarajan13}. 
The estimator is closely related, both in concept and form, to the well-known CMB lensing estimators \cite[e.g.,][]{seljak99, hu01b}. 
\cite{dvorkin09} estimate that a future CMB polarization satellite, in addition to mapping the lensing B-modes, might measure the duration of reionization and mean bubble size  at the 10\% level. 
Such a measurement could tell us about how quickly reionization occurred and what kind of sources were responsible for reionization. 

The main observational challenge is the faintness of the signal. 
Compared to the recently detected lensing B-modes \cite{hanson13, polarbear14b}, the B-mode power induced by the variations in optical depth is lower by a factor of 10-100 depending on angular scale (see Figure 3 in \cite{dvorkin09}). 
Still given the high signal-to-noise ($> 100\,\sigma$) expected for lensing B-modes from CMB experiments under construction  (e.g., \advact{}, \sptnew{} \cite{benson14}, \simons), and the substantially better mapping speeds that would be achieved by the proposed stage IV ground-based experiments or the next satellite experiment, the reionization signal should eventually be detectable at high signal-to-noise. 
While unlikely to be competitive with 21\,cm surveys on this timescale, mapping the variations in optical depth across the sky with the CMB would be an independent test of reionization scenarios.

\section{Conclusions}
\label{sec:conclusion}

The CMB has yielded two major clues into cosmic reionization to date: the optical depth and kSZ power. 
In this work, we have discussed the theory and observations for the two observables, as well as mentioning the possibility of constraining the temperature of the intergalactic medium during reionization and mapping variations in the optical depth across the sky. 

First and most robustly, measurements of large-scale CMB polarization anisotropy can be used to determine the optical depth of the Universe due to Thomson scattering. 
The \wmap{} 9-year results favor $\tau = 0.089 \pm 0.014$ \cite{bennett13}. 
The optical depth  due to Thomson scattering depends on the column depth of free electrons, and is roughly proportional to the redshift at which the Universe is 50\% ionized. 
The \wmap{} measurement suggests the midpoint of reionization is around $z = 10.6 \pm 1.1$. 
In the near future, the \planck{} satellite is expected to publish an independent measurement of the optical depth with a factor of three reduction in errors. 

Second, multifrequency measurements of small-scale (few arcminutes) CMB temperature anisotropy have begun to set interesting upper limits on the kSZ power. 
When combined with simulations of the homogenous kSZ power from the fully ionized Universe, the upper limits on the total kSZ power suggest reionization was fairly rapid. 
\cite{george14} find a 95\% CL upper limit on redshift interval in which the Universe transitioned from  20\% to 99\% ionized to be $\Delta z < 5.4$. 
Measurements of the  kSZ power should be  substantially improved by experiments under construction or planned, with first light instruments on \ccat{} expected to improve upon current measurements by a factor of approximately 16.

CMB observations have been among the first direct observational probes of the EoR, and will continue to provide new and independent tests of how cosmic reionization occurred  going forward. 
The CMB is complementary to planned 21\,cm experiments as the CMB observables probe the ionized rather then neutral gas. 
The main insights from the CMB so far have been on the timing of reionization, with the optical depth setting when cosmic reionization occurred and the limits on kSZ power constraining the duration of reionization. 
In the future, measurements of the CMB may also test of the nature of the first objects, with the shape of the kSZ power spectrum and variations in optical depth across the sky probing bubble sizes and spectral distortions probing the temperature of the intergalactic medium.

\comment{
%
\begin{figure}[b]
\sidecaption
\includegraphics[scale=.65]{figure}
%
%
\caption{If the width of the figure is less than 7.8 cm use the \texttt{sidecaption} command to flush the caption on the left side of the page. If the figure is positioned at the top of the page, align the sidecaption with the top of the figure -- to achieve this you simply need to use the optional argument \texttt{[t]} with the \texttt{sidecaption} command}
\label{fig:1}       
\end{figure}
}

\begin{acknowledgement}
We thank Oliver Zahn for useful discussions and for creating Figure~\ref{fig:ccat}. 
We are grateful to Elizabeth George, Kyle Story, and Andrei Mesinger for valuable feedback. 
We acknowledge the use of the Legacy Archive for Microwave Background Data Analysis (LAMBDA). 
Support for LAMBDA is provided by the NASA Office of Space Science. 
\end{acknowledgement}
%
%
%
\comment{When placed at the end of a chapter or contribution (as opposed to at the end of the book), the numbering of tables, figures, and equations in the appendix section continues on from that in the main text. Hence please \textit{do not} use the \verb|appendix| command when writing an appendix at the end of your chapter or contribution. If there is only one the appendix is designated ``Appendix'', or ``Appendix 1'', or ``Appendix 2'', etc. if there is more than one.}

\bibliographystyle{apj}
\bibliography{ms}{}

\end{document}